\documentclass[AMA,STIX1COL,doublespace]{WileyNJD-v2}
\usepackage{moreverb}

\usepackage{setspace}
\usepackage{hyperref}
\hypersetup{
	colorlinks   = true, %Colours links instead of ugly boxes
	urlcolor     = blue, %Colour for external hyperlinks
	linkcolor    = blue, %Colour of internal links
	citecolor   = blue %Colour of citations
}

\newcommand\BibTeX{{\rmfamily B\kern-.05em \textsc{i\kern-.025em b}\kern-.08em
T\kern-.1667em\lower.7ex\hbox{E}\kern-.125emX}}

\articletype{Research Article}%

\received{<day> <Month>, <year>}
\revised{<day> <Month>, <year>}
\accepted{<day> <Month>, <year>}

\begin{document}

\title{Improving the Performance of Bayesian Logistic Regression Model with Overdose Control in Oncology Dose-Finding Studies}

\author[1]{Hongtao Zhang*}

\author[1]{Alan Y Chiang}

\author[2]{Jixian Wang}

\authormark{Zhang, H. \textsc{et al}}

\address[1]{\orgdiv{Global Biometrics and Data Sciences}, \orgname{Bristol Myers Squibb}, \orgaddress{\state{Berkeley Heights, New Jersey}, \country{USA}}}

\address[2]{\orgdiv{Global Biometrics and Data Sciences}, \orgname{Bristol Myers Squibb}, \orgaddress{\state{Boudry}, \country{Switzerland}}}

%\address[2]{\orgdiv{Org Division}, \orgname{Org name}, \orgaddress{\state{State name}, \country{Country name}}}

\corres{*Hongtao Zhang, \email{squallteo@gmail.com}}

%\presentaddress{300 Connell Drive, Berkeley Heights, NJ 07922, USA}

\abstract[Abstract]{An accurately identified maximum tolerated dose (MTD) serves as the cornerstone of successful subsequent phases in oncology drug development. Bayesian logistic regression model (BLRM) is a popular and versatile model-based dose-finding design. However, BLRM with original overdose control strategy has been reported to be safe but ``excessively conservative''. In this manuscript, we investigate the reason for conservativeness and point out that a major reason could be the lack of appropriate underdose control. We propose designs that balance overdose and underdose control to improve the performance over original BLRM. Simulation results reveal that the new designs have better accuracy and treat more patients at MTD.}


%\jnlcitation{\cname{%
%\author{Hongtao Zhang},
%and
%\author{Alan Y Chiang} (\cyear{2021+}), 
%\ctitle{New Overdose Control Designs for Bayesian Logistic Regression Model in Oncology Dose-Finding}, \cjournal{Statistics in Medicine}, \cvol{xxx}.}
\maketitle

%\footnotetext{\textbf{Abbreviations:} ANA, anti-nuclear antibodies; APC, antigen-presenting cells; IRF, interferon regulatory factor}

\section{Introduction}
One primary objective of phase I oncology dose-finding trials is to identify the maximum tolerated dose (MTD) of the drug candidate to inform the dose level(s) to be investigated in subsequent phases of development. In most cases, the data used to determine MTD is a binary scale based on the presence or absence of dose-limiting toxicity (DLT) within a certain period (e.g., 28 days), where DLT is usually defined as treatment related non-hematological toxicity of Grade 3 or higher, or treatment related hematological toxicity of Grade 4 or higher based on the National Cancer Institute Common Terminology Criteria for Adverse Events (CTCAE). Cohorts of patients are enrolled into provisional dose levels, usually starting from the lowest or second lowest one, in a sequential and adaptive manner. Upon observing the DLT data from the latest cohort of patients, a recommendation is rendered for the dose level to be assigned to the next cohort of patients based on a certain dose-finding design. This process is repeated until the total sample size is exhausted or certain pre-specified early-stopping rules are met. 

Oncology dose-finding is an extremely active area for innovative researches in study designs. Recent developments in dose-finding designs have largely shifted away from the the rule-based 3+3 paradigm. One popular class is the model-assisted designs as represented by Bayesian optimal interval (BOIN \cite{LiuYuan_2015BOIN}) and modified toxicity probability interval (mTPI \cite{JiEtal_2010mTPI,GuoEtal_2017mTPI2}) designs. One advantage of BOIN and mTPI designs is the ease of implementation for clinical practitioners as the actions (escalate/stay/deescalate) corresponding to observed data can be summarized through a pre-planned table, making the dose recommendation process very transparent and easy to follow. Another class is the model-based designs which explicitly specify an underlying parametric dose-toxicity relationship. Continual reassessment method (CRM \cite{OQuigleyEtal_1990}) is the first model-based design to find the MTD that is closest to a target toxicity level. Similar to CRM, Bayesian logistic regression model (BLRM \cite{NeuenschwanderEtal_2008BLRM}) relies on the two-parameter logistic model to identify the MTD that falls into a predefined target toxicity interval for DLT rate, e.g., (0.16, 0.33). Compared to model-assisted designs, CRM and BLRM involve Bayesian regression of the underlying parametric models which is usually more computationally intensive, and the aforementioned decision table is unavailable. However, model-based designs, driven from modeling dose-toxicity paradigm, have shown successes in more sophisticated extensions. For example, extensions \cite{Neuenschwander_2015BLRMCovar} of BLRM have been proposed for situations when there is additional covariate information or drug combination, potentially with drug-drug interaction. Some dose-finding trials simultaneously enroll patients into multiple strata, for example, indications and populations. With BLRM, robust information sharing across homogeneous strata is made possible via exchangeability-non-exchangeability (EXNEX) model \cite{NeuenschwanderEtal_2016} or semiparametric Dirichlet process prior \cite{LiEtal_2020BSD}. Pharmacokinetic \cite{CotterillEtal_2015} or animal \cite{ZhengEtal_2020} data can also be incorporated in BLRM. 

In phase 1 oncology dose-finding studies, some patients may be unavoidably assigned to dose levels that are below or above the MTD, resulting in underdosing and overdosing respectively. While both are unfavorable outcomes, overdosing patients is generally considered more of patient safety concern. Therefore, most dose-finding designs have mechanisms that control the probability of overdosing. The concept of escalation with overdose control (EWOC) was first introduced by Babb and colleagues \cite{BabbEtal_1998CRMEWOC} based on CRM. The idea is to recommend a dose level that is lower than the median of posterior MTD distribution, resulting in a more conservative design than CRM \cite{LinYin_2017EWOC, ChuEtal_2009}. The key concept underlying EWOC is that one can select dose levels for use in a phase I trial so that the predicted proportion of patients who receive an overdose is equal to a specified feasibility bound. With a similar framework, EWOC can be extended to be derived from toxicity-dependent feasibility bounds \cite{Wheeler_2017TDEWOC} or in a non-parametric manner \cite{LinYin_2017EWOC}. In BLRM, overdose control is done by limiting the probability of overdosing below a certain threshold, e.g. 0.30. Popular model-assisted designs like BOIN and mTPI(-2) also account for overdosing in the process of deriving the optimal designs. We defer a more detailed review of overdose control mechanisms to Section 2. 

Simulation studies comparing the operating characteristics across various dose-finding designs are aplenty \cite{AnanthakrishnanEtal_2017,ZhouEtal_2018DFCompare,ZhouEtal_2018DFCompare_NoBLRM}. In particular, BLRM with overdose control has been reported to be ``excessively conservative'' \cite{ZhouEtal_2018DFCompare}. This observation has also been made in real dose-finding studies with BLRM. Table \ref{tab:data_scenario} illustrates a hypothetical data scenario in which BLRM is frustratingly conservative. The dose escalation recommendation is made by BLRM with (0.16, 0.33) target toxicity interval and overdosing control bound of 0.25 \cite{NeuenschwanderEtal_2008BLRM}. Three patients are enrolled in each of the first four provisional doses and no DLT has been observed. However, despite the fact that the current dose (100mg) is clearly underdosing, BLRM doesn't recommend escalation to the next level (200mg) because the probability of overdosing at 200mg is slightly higher than the overdosing control bound 0.25. 

\begin{table}[htbp]
	\centering
	\caption{A Hypothetical Data Scenario}
	\begin{tabular}{cccccc}
		Dose (mg) & \#DLT & \#Patient & P(Under) & P(Target) & P(Over) \\
		\hline
	    10    & 0     & 3     & 0.998 & 0.002 & 0.000 \\
		25    & 0     & 3     & 0.992 & 0.008 & 0.000 \\
		50    & 0     & 3     & 0.972 & 0.027 & 0.001 \\
		100   & 0     & 3     & 0.777 & 0.186 & 0.037 \\
		200   & 0     & 0     & 0.510 & 0.200 & \textbf{0.290} \\
		400   & 0     & 0     & 0.344 & 0.190 & 0.467 \\
		800   & 0     & 0     & 0.239 & 0.167 & 0.595 \\
	\end{tabular}%
	\label{tab:data_scenario}%
\end{table}%

In this manuscript, we focus on improving the performance of the original overdose control strategy in BLRM. In Section 2, we first align how the target toxicity level/interval is defined across BLRM and other designs. We then review the overdose control mechanisms in popular dose-finding designs, point out the gap between BLRM and others, and propose more balanced overdose control designs for BLRM. Simulation results are presented in Section 3 comparing new designs with the original BLRM. Finally, Section 4 provides concluding remarks about future research and other practical considerations. 

\section{Method}
\subsection{Target Toxicity Level/Interval}
The concept of target toxicity level (TTL), denoted by the scalar $\phi$, was introduced in CRM \cite{OQuigleyEtal_1990}. The goal of CRM is to select an MTD that has the estimated DLT rate as close to $\phi$ as possible. However, the sample sizes of phase 1 dose-finding studies are generally small or moderate, rendering it impractical to pinpoint the MTD with great precision. In other words, it is difficult to find an MTD whose DLT rate is precisely or very close to $\phi$. Acknowledging this, most contemporary dose-finding designs are interval-based, which allow some level of deviation around the TTL. In interval-based designs, an target toxicity interval (TTI) is determined and the MTD should have a DLT rate falling into the TTI. In BOIN design\cite{LiuYuan_2015BOIN}, such deviation is multiplicative in the form of ($m_L\phi, m_U\phi$). The recommended multipliers are $m_L=0.6$ and $m_U=1.4$ for lower and upper bounds, respectively. For example, with a TTL of $\phi=0.25$, the TTI in BOIN design with recommended multipliers is (0.15, 0.35). In mTPI designs\cite{JiEtal_2010mTPI,GuoEtal_2017mTPI2}, TTI (termed equivalence interval) is constructed using additive deviation ($\phi - \epsilon_1, \phi + \epsilon_2$), where common choices of $\epsilon$'s are 0.05 and 0.1. By using $\epsilon_1 = \epsilon_2 = 0.1$, the same TTI as in the BOIN example can be obtained with $\phi=0.25$. 

In contrast, while BLRM also adopts a TTI, the concept of TTL did not apply when BLRM was first introduced\cite{NeuenschwanderEtal_2008BLRM}. Therefore, a common misconception in practice when interpreting BLRM is that it sets the upper bound of TTI to the TTL, for instance, $\phi=0.35$ when TTI is (0.15, 0.35). This is not necessarily true because TTI generally covers DLT rates that are higher than the TTL $\phi$. Apparently, the concept of TTL can be introduced into BLRM without any modification, and it behooves to align the process of determining the TTI in BLRM with other interval-based designs such as BOIN and mTPI. 

\subsection{Overdose Control Mechanisms in Dose-Finding Designs}
Given the focus on safety in oncology dose-finding studies, the concept of escalation with overdose control (EWOC) was first introduced by Babb and colleagues\cite{BabbEtal_1998CRMEWOC} based on CRM. Their method assumes that the DLT rate at dose $d$ is $F(\alpha + \beta d)$ with linear predictor $\alpha + \beta d$, where $F$ is a specified distribution function. The marginal posterior distribution of MTD, denoted by $\pi(\cdot)$, can then be derived. Upon observing the toxicity data of the latest patient, the recommended dose $d'$ for the next patient is the $\alpha_f$-th percentile of the posterior MTD distribution, that is, $d' = \pi^{-1}(\alpha_f)$. The fraction $\alpha_f$ is known as the feasibility bound. It was pointed out\cite{LinYin_2017EWOC,ChuEtal_2009} that CRM tends to assign patients to the median of the posterior MTD distribution, in which case $\alpha_f=0.5$. With EWOC, the authors suggested using a feasibility bound $\alpha_f < 0.5$, resulting in a more conservative design than CRM. Equivalently, the recommended dose $d'$ for the next patient minimizes the risk with respect to an asymmetric loss function
\begin{singlespace}
	\begin{equation*}
		l_{\alpha_f}(d', d_\phi) = 
		\begin{cases}
			\alpha_f (d_\phi - d') & \text{, if $d' \leq d_\phi$,} \\
			(1-\alpha_f) (d' - d_\phi) & \text{, if $d' > d_\phi$.}
		\end{cases}
	\end{equation*}
\end{singlespace}

Here $d_\phi$ is the dose level corresponding to the TTL ($\phi$). When $\alpha_f < 0.5$, it is implied that the loss (or penalty) is lower for underdosing than overdosing, corresponding to the same deviation from TTL, or $|d'-d_\phi|$. For example, the loss for overdosing is three times of that for underdosing with $\alpha_f = 0.25$ for the same $|d'-d_\phi|$. It is worth noting that the concept of TTI was not used in EWOC. Subsequent researches proposed that the feasibility bound can gradually increase to 0.5 with patient accrual\cite{BabbRogatko_2001} or be determined based on the number of DLTs observed\cite{Wheeler_2017TDEWOC}. Recently, the non-parametric overdose control design\cite{LinYin_2017EWOC} (NOC) extends the EWOC framework by eliminating the need to specify the distribution function $F$. A symmetric TTI centered at TTL $\phi$ was utilized in this method to define MTD. The probability of each dose being the MTD is derived in a non-parametric manner. The recommended dose for the next patient minimizes an asymmetric loss function that is very similar to that in EWOC, per the feasibility bound. In other words, both underdosing and overdosing are penalized, albeit differently. This method also has a dosing-switch rule that overrides the overdose control. Specifically, if there is overwhelming evidence that a certain dose level is the MTD, then the next patient will be assigned to it regardless whether overdose control requirement is satisfied. 

We now focus on two commonly used classes of interval-based designs, namely BOIN and mTPI designs. Denote $[a, b]$ the TTI and actions $\{\mathcal{E, S, D}\}$ for escalate, stay and deescalate respectively. Depending on the placement of DLT rate $p_i$ at the current dose level $i$ with respect to TTI, the actions can be classified as correct or incorrect, as shown in Table \ref{tab:actions}. 
\begin{table}[htbp]
	\centering
	\caption{Correct and Incorrect Actions}
	\label{tab:actions}
	\begin{tabular}{ccc}
		Scenario & Correct Action & Incorrect Action \\
		\hline
		$p_i \in [0, a]$ & $\mathcal{E}$ &  $\bar{\mathcal{E}}=\mathcal{\{S, D\}}$ \\
		$p_i \in (a, b)$ & $\mathcal{S}$ &  $\bar{\mathcal{S}}=\mathcal{\{E, D\}}$ \\
		$p_i \in [b, 1]$ & $\mathcal{D}$ &  $\bar{\mathcal{D}}=\mathcal{\{E, S\}}$ \\
	\end{tabular}
\end{table}%

In BOIN \cite{LiuYuan_2015BOIN} design, three point hypotheses are defined
\begin{equation*}
	H_{0i}: p_i = \phi,\quad H_{1i}: p_i = a,\quad H_{2i}: p_i = b,
\end{equation*}
which indicate that the current dose level $i$ is at/below/above MTD so that the correct action should be $\mathcal{S}/\mathcal{E}/\mathcal{D}$, respectively. The probability of making an incorrect decision, or decision error rate, at each dose assignment is given by
\begin{equation*}
	\alpha(\lambda_{1i}, \lambda_{2i}) = pr(H_{0i})pr(\bar{\mathcal{S}}|H_{0i}) + pr(H_{1i})pr(\bar{\mathcal{E}}|H_{1i}) + pr(H_{2i})pr(\bar{\mathcal{D}}|H_{2i}),
\end{equation*}
where $\lambda_{1i}$ and $\lambda_{2i}$ are dose escalation and deescalation boundaries. The optimal $\lambda_{1i}$ and $\lambda_{2i}$ minimize the decision error rate and are determined by sample size, $\phi$, TTI and the prior. It is clear that all incorrect decisions contribute to, or are penalized in, the decision error rate. For example, underdosing (failure to escalate) is penalized when the current dose is below MTD ($H_{1i}$). The BOIN design also has a dose elimination rule to terminate overly toxic doses. 

In the case of mTPI design\cite{JiEtal_2010mTPI}, the loss functions $L(\mathcal{D},p_i)$, $L(\mathcal{S},p_i)$ and $L(\mathcal{E},p_i)$ for taking respective actions are explicitly defined for dose level $i$. The loss is positive unless the correct action is taken at dose level $i$ in which case there is zero loss. The dose-finding algorithm minimizes the posterior expected loss. In addition, mTPI design also has two safety rules. One terminates the trial early due to excessive toxicity, while the other one eliminates overly toxic doses. The mTPI-2 design\cite{GuoEtal_2017mTPI2} considered a 0-1 loss function. The loss is 0 if the correct action is taken, and 1 otherwise. Similar to mTPI, the mTPI-2 design also minimizes the posterior expected loss based on the 0-1 loss function. 

\subsection{New Overdose Control Designs for BLRM}
The overdose control mechanisms reviewed in Section 2.2 have one feature in common: all incorrect decisions including overdosing and underdosing are penalized in or contribute to the loss function used to derive the dose-finding algorithm. In EWOC and NOC, overdosing is penalized more heavily than underdosing, as the former is generally deemed a more unfavorable outcome. In BOIN and mTPI-2, the same penalty is imposed on all incorrect decisions. The mTPI design, on the other hand, allows flexible loss functions so that overdosing and underdosing may or may not be penalized differently. 

In contrast, the existing overdose control strategy in BLRM\cite{NeuenschwanderEtal_2008BLRM} only limits the probability overdosing and leaves underdosing unpunished. To see this, consider the BLRM with DLT rate $p_i$ at dose $d_i$ for $i = 1,\dots,K$
\begin{equation}\label{eqn:blrm}
	logit(p_i) = \log{\alpha} + \beta\log{\frac{d_i}{d_R}},
\end{equation}
where $\alpha, \beta > 0$ and $d_R$ is the reference dose level allowing for the interpretation of $\alpha$ as the odds of a DLT at $d_R$. A bivariate normal prior is assigned to $(\log\alpha, \log\beta)'$. The domain of DLT rate $[0, 1]$ is partitioned into non-overlapping toxicity intervals which, in the order of increasing toxicity, are underdosing, target toxicity, excessive toxicity and unacceptable toxicity. In practice, the last two overdosing categories may be combined and we assume this is the case hereafter. Upon observing the toxicity data from latest cohort of patient(s), the cumulative data are fitted using BLRM and the probabilities of underdosing/on-target/overdosing are calculated for each dose level. The recommended dose level is the one with highest probability of being in the TTI. With overdose control, the probability of overdosing at the recommended dose level should not exceed a pre-specified bound which is usually 0.25 or 0.3. There is no restriction on the probability of underdosing which may result in frustrating situations like the hypothetical scenario in Table \ref{tab:data_scenario}. Due to safety considerations, skipping dose levels during dose escalations is not recommended. 

We believe the lack of ``underdosing control'' is the main reason that BLRM with overdose control is found to be conservative. To overcome this, we propose the following new overdosing control designs for BLRM that also account for underdosing. Each of the new designs has an add-on rule that overrides the existing overdose control mechanism in BLRM. The add-on rule is assessed first in the new designs unless the study is currently at the highest provisional dose level. If met, the recommended action is escalation by one dose level, overriding the existing overdose control based on probability of overdosing. Otherwise, the dose recommendation is rendered in the same manner as described above. The advantage of using the add-on rule is that the probability of underdosing can be factored in without fundamentally altering the dose recommendation framework in BLRM. 

\textbf{Design 1}: We introduce the feasibility bound $\alpha_f < 0.5$. At current dose $d_i$, the add-on rule to trigger escalation is met if 
\begin{equation}\label{eqn:design1}
	\alpha_f P_i(Under) > (1-\alpha_f) P_{i}(Over),
\end{equation}
where $P_i(Under)$ and $P_i(Over)$ are probabilities that the DLT rate at current dose $d_i$ falls into underdosing and overdosing intervals respectively. This design is reminiscent of the original EWOC method\cite{BabbEtal_1998CRMEWOC} in the sense that the feasibility bound has a similar interpretation. 
%The term on the left-hand side of \eqref{eqn:design1} can be considered the loss of underdosing, whereas the term on the right-hand side the loss of overdosing. When the former exceeds the latter, it is intuitive to recommend an escalation action. 

\textbf{Design 2}: The add-on rule \eqref{eqn:design1} is assessed using only the probabilities at current dose level $d_i$. Since BLRM specifies a monotonically increasing dose-toxicity relationship, the dose recommendation also depends on the relative strength of current dose with respect to the next dose, $r_i = d_{i+1}/d_i, \forall i < K$. For example, while BLRM may not recommend an escalation when $r_i = 2$, it may be permitted when $r_i = 1.5$. Therefore, we compare the probability of underdosing at current dose level with the probability of overdosing at next dose level. Specifically, the add-on rule is met if
\begin{equation}\label{eqn:design2}
	P_i(Under) > g(r_i) \cdot P_{i+1}(Over),
\end{equation}
where $g(\cdot)$ is a non-decreasing function of $r_i$. A non-decreasing $g(\cdot)$ guarantees that the add-on rule is less likely to be met with a larger $r_i$, or equivalently, a larger dose increment. An apparent choice for $g(\cdot)$ is the identity function $g(r_i) = r_i$. 

The add-on rules in designs 1 and 2 are based on the probabilities that the DLT rate falls into underdosing and overdosing intervals, respectively. The width of underdosing/overdosing interval clearly has an impact on these probabilities. With the same observed data, the wider the interval, the higher the probability of falling into it. Hence we may standardize the probabilities by dividing them by the width of the corresponding intervals which, in effect, yields the unit probability mass (UPM) used in the mTPI design\cite{JiEtal_2010mTPI}. Specifically, we have $\mathcal{U}_i(Under) = P_i(Under)/a$ and $\mathcal{U}_i(Over) = P_i(Over)/(1-b)$, where $a, b$ are defined in Table \ref{tab:actions}. We propose designs 3 and 4 with UPM-based add-on rules. 

\textbf{Design 3}: This design is similar to design 1, except that the add-on rule is met if
\begin{equation*}
	\alpha_f \mathcal{U}_i(Under) > (1-\alpha_f) \mathcal{U}_{i}(Over).
\end{equation*}

\textbf{Design 4}: This design is similar to design 2, except that the add-on rule is met if
\begin{equation*}
	\mathcal{U}_i(Under) > g(r_i) \cdot \mathcal{U}_{i+1}(Over).
\end{equation*}

One caveat of UPM is that it has less clear interpretation compared to the interval probabilities. Therefore, in designs 3 and 4, it is intended that UPMs are only used in the add-on rules in which the UPMs of underdosing and overdosing are compared. Other elements in BLRM such as overdose control and dose recommendation are still based on interval probabilities. 

\section{Numerical Studies}
We conduct simulation studies to compare the operating characteristics of new overdose control designs with the original BLRM. The first part of simulation uses three fixed dose-toxicity scenarios whose shapes are frequently encountered in practice. In the second part, we use randomly generated dose-toxicity scenarios to ensure that the trends observed in the first part are not due to cherry-picking of scenarios. R programs are available on GitHub: \textcolor{blue}{(link will be shared upon manuscript acceptance)}. 

Simulations across designs and fixed/random scenarios share some common specifications. Each simulated trial has a maximum of 45 patients that are enrolled in cohorts of size of 3 patients. Across the board, the provisional dose levels are 10, 25, 50, 100, 200, 400 and 800 in milligrams. The same BLRM \eqref{eqn:blrm} is fitted with reference dose 100mg and a weakly-informative bivariate normal prior for the regression parameters
\begin{equation*}
	\bigg[ \begin{array}{cc}
		\log\alpha \\ \log\beta
	\end{array} \bigg] \sim N \bigg( \bigg[ \begin{array}{cc}
	-0.693 \\ 0
\end{array} \bigg], \bigg[ \begin{array}{cc}
4 & 0 \\ 0 & 1
\end{array} \bigg] \bigg). 
\end{equation*}
Echoing with the discussions in Section 2.1, we set TTL at $\phi = 0.25$ and consider two TTIs centered at it: (0.20, 0.30) and (0.16, 0.33). The limit for the probability of overdosing interval, as in original overdosing control strategy, is 0.3. A dose level can be declared as the MTD if all following criteria are met: 
\begin{itemize}
	\item BLRM recommends to stay at this dose level;
	\item A minimum of six patients have been enrolled at this dose level;
	\item For TTI (0.16, 0.33), the probability of target toxicity is at least 0.5. For TTI (0.20, 0.30), the probability of target toxicity is at least 0.4, acknowledging a narrower TTI. 
\end{itemize}
Upon updating the BLRM with data from the latest cohort of patients, we may stop the trial early in either of the following cases: 
\begin{itemize}
	\item All doses are deemed overly toxic. In other words, the probability of overdosing interval is above 0.3 for all doses. In this case, the MTD is below the lowest provisional dose level; 
	\item An MTD can be declared based on the available data. 
\end{itemize}
If the trial is neither stopped early nor at the highest provisional dose level, respective add-on rules in new designs will be evaluated to determine whether the dose should be escalated by one level. The dose recommendation will be rendered by existing dose recommendation framework in BLRM if the add-on rule is not triggered. Specifically, the recommended action will be $\mathcal{E}/\mathcal{S}/\mathcal{D}$ when the dose level with highest probability of target toxicity is above/at/below the current dose level, respectively. It is also required that the probability of overdosing interval at such dose level must not exceed 0.3. For practical considerations, we also require that skipping dose levels is not permitted. Finally, it is possible that an MTD still cannot be declared after we reach the maximum sample size. In this case, the trial is terminated and an MTD is not found. 

The original overdose control and new designs in BLRM are evaluated in simulation with the following specifications: 
\begin{singlespace}
	\begin{itemize}
		\item Original overdose control at 0.3;
		\item Design 1 with feasibility bound $\alpha_f = 0.25$ and overdose control at 0.3;
		\item Design 2 with $g(r_i) = r_i$ and overdose control at 0.3;
		\item Design 3 with feasibility bound $\alpha_f = 0.25$ and overdose control at 0.3;
		\item Design 4 with $g(r_i) = r_i$ and overdose control at 0.3.
	\end{itemize}
\end{singlespace}

\subsection{Simulation with Fixed Dose-Toxicity Scenarios}
\begin{figure}
	\centering
	\includegraphics[width=0.9\textwidth, keepaspectratio]{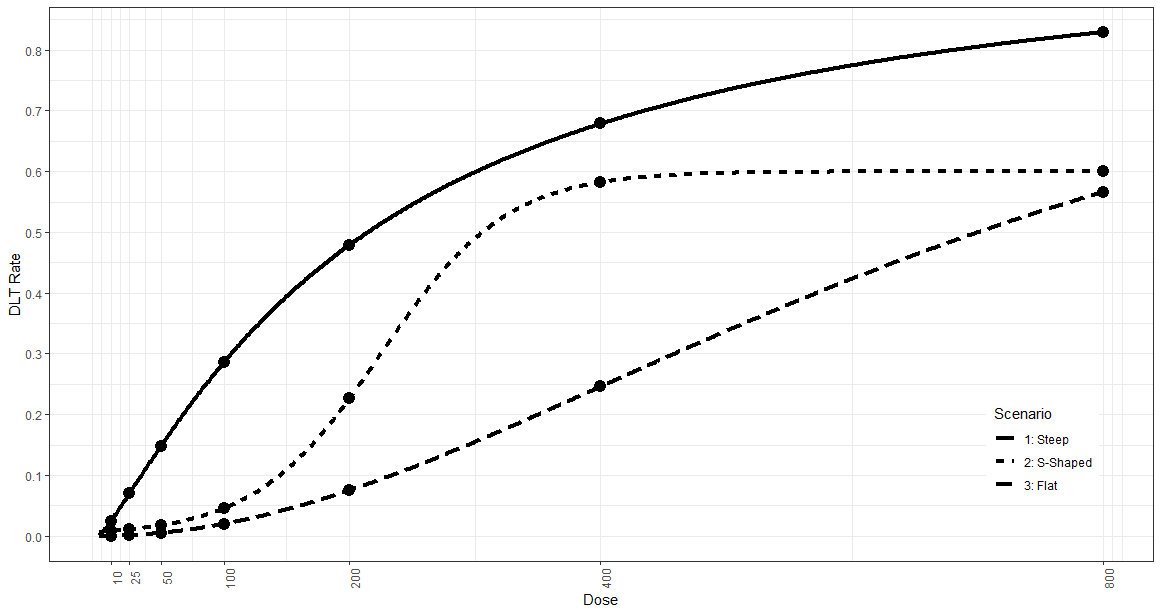}
	\caption{Fixed Dose-Toxicity Scenarios}
	\label{fig:fixed_scenario}
\end{figure}

We first present the simulation results in which three fixed dose-toxicity scenarios, namely, steep, S-shaped and flat dose-toxicity curves are considered. Each shape is generated from a parametric model (Table \ref{tab:paramodel} and Figure \ref{fig:fixed_scenario}). The DLT rates at provisional dose levels are presented in Table \ref{tab:fixed_scenario}. Each design is evaluated in 1,000 simulations in each dose-toxicity scenario. 

\begin{table}[h]
	\centering
	\caption{Parametric Models in Fixed Dose-Toxicity Scenarios}
	\begin{tabular}{cc}
		Shape & $p_i$ \\
		\hline
		Steep & $\{ 1 + \exp\{-0.916 + 1.2*\log(d_i/100)\} \}^{-1}$ \\
		S-shaped & $0.6 * \{ 1 + \exp\{-0.02*(d_i - 225)\} \}^{-1}$ \\
		Flat & $\{ 1 + \exp\{-2*\log(d_i/700)\} \}^{-1}$
	\end{tabular}
	\label{tab:paramodel}
\end{table}

\begin{table}[htbp]
	\centering
	\caption{DLT Rates in Fixed Scenario Simulations}
	\begin{tabular}{rrrr}
		\multicolumn{1}{c}{\multirow{2}[0]{*}{Dose}} & \multicolumn{3}{c}{Scenario} \\
		& \multicolumn{1}{l}{Steep} & \multicolumn{1}{l}{S-shaped} & \multicolumn{1}{l}{Flat} \\
		\hline
		10    & 0.025 & 0.008 & 0.000 \\
		25    & 0.070 & 0.011 & 0.001 \\
		50    & 0.148 & 0.018 & 0.005 \\
		100   & 0.286 & 0.046 & 0.020 \\
		200   & 0.479 & 0.227 & 0.076 \\
		400   & 0.679 & 0.582 & 0.246 \\
		800   & 0.829 & 0.600 & 0.566 \\
	\end{tabular}%
	\label{tab:fixed_scenario}%
\end{table}%

\begin{figure}
	\centering
	\includegraphics[width=\textwidth, height=\textheight, keepaspectratio]{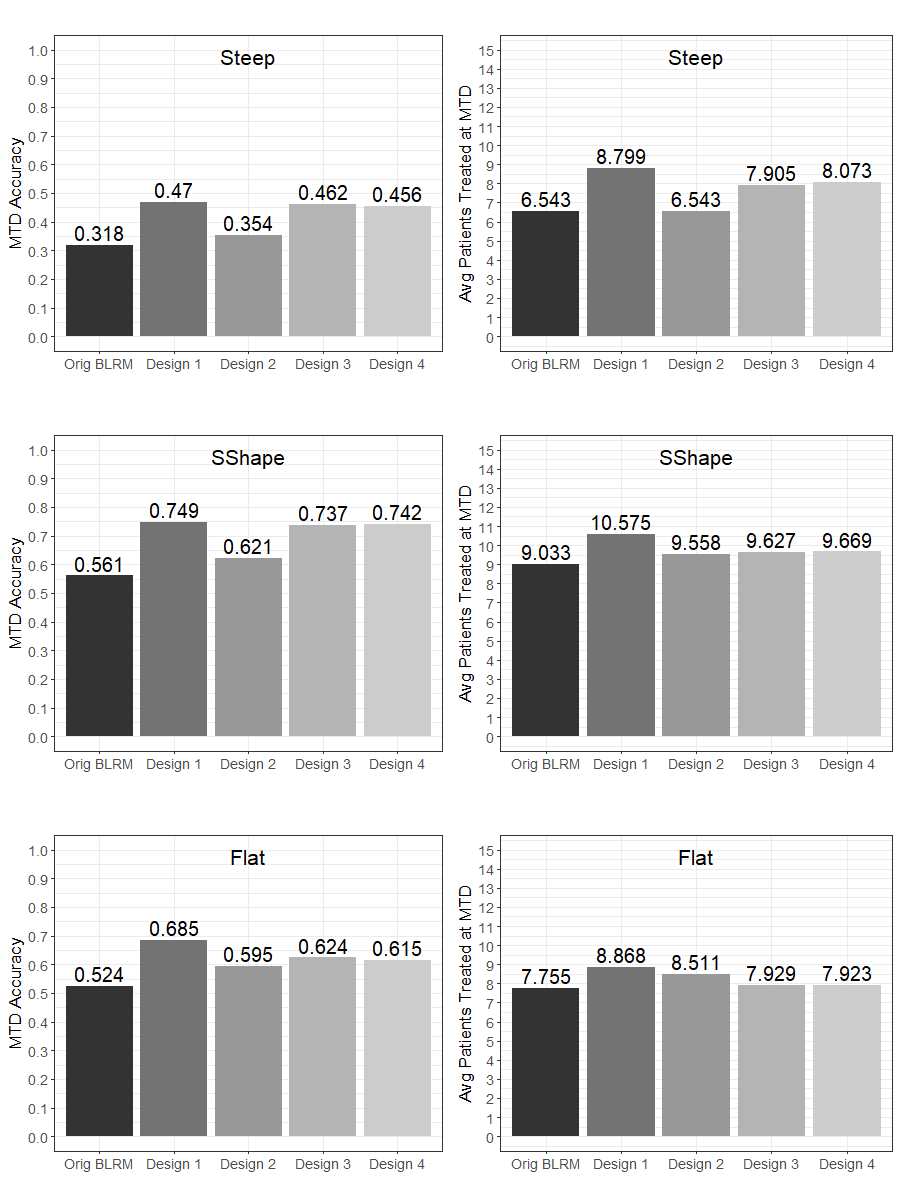}
	\caption{Key Simulation Results with TTI (0.16, 0.33)}
	\label{fig:16_33}
\end{figure}

Figure \ref{fig:16_33} presents two most important operating characteristics: the frequency of correctly identified MTD (left column) and the average number of patients treated at MTD (right column), with TTI (0.16, 0.33). It is clear that all new overdose control designs outperform the strategy in original BLRM in terms of having higher accuracy of identifying MTD and assigning more patients to the MTD. In the results shown, design 1 has the best performance, while design 2 behaves most similar with the original BLRM. Designs 3 and 4 yield very similar operating characteristics, which is generally the case across all simulations. Complete simulation results are provided in Appendix A. Table \ref{tab:16_33_sshaped} shows full simulation results for TTI (0.16, 0.33) with the S-shaped dose-toxicity curve. The row in boldface indicates the true MTD in this scenario. In addition to the aforementioned observations, the new designs on average require a smaller overall sample size than original BLRM. On the other hand, they also tend to assign slightly more patients to dose levels higher than the true MTD, and have slightly higher observed overall DLT rate. However, these are reasonable compromises considering the promising improvements in other metrics. In other simulations with the same TTI, the new designs also perform better than original BLRM and we observe similar trends in operating characteristics. 

\begin{table}[htbp]
	\centering
	\caption{Results: TTI (0.16, 0.33), S-shaped Curve}
	\begin{tabular}{c|cc|cc|cc|cc|cc}
		& \multicolumn{2}{c}{Original BLRM} & \multicolumn{2}{c}{Design 1} & \multicolumn{2}{c}{Design 2} & \multicolumn{2}{c}{Design 3} & \multicolumn{2}{c}{Design 4} \\
		MTD   & Frequency$^\dagger$ & N$^\ddagger$     & Frequency & N     & Frequency & N     & Frequency & N     & Frequency & N \\
		\hline
		AllToxic$^1$ & 0.017 & -     & 0.017 & -     & 0.017 & -     & 0.017 & -     & 0.017 & - \\
		10    & 0.000 & 3.01  & 0.000 & 3.01  & 0.000 & 3.01  & 0.000 & 3.00  & 0.000 & 3.00 \\
		25    & 0.000 & 3.07  & 0.000 & 3.03  & 0.000 & 3.07  & 0.000 & 2.97  & 0.000 & 2.97 \\
		50    & 0.000 & 3.74  & 0.000 & 3.11  & 0.000 & 3.70  & 0.001 & 3.02  & 0.001 & 3.03 \\
		100   & 0.047 & 14.30 & 0.051 & 7.91  & 0.045 & 12.08 & 0.060 & 6.62  & 0.057 & 6.70 \\
		\textbf{200} & \textbf{0.561} & \textbf{9.03} & \textbf{0.749} & \textbf{10.58} & \textbf{0.621} & \textbf{9.56} & \textbf{0.737} & \textbf{9.63} & \textbf{0.742} & \textbf{9.67} \\
		400   & 0.028 & 2.36  & 0.044 & 3.73  & 0.026 & 2.53  & 0.052 & 3.99  & 0.053 & 3.88 \\
		800   & 0.002 & 0.22  & 0.002 & 0.30  & 0.001 & 0.21  & 0.004 & 0.71  & 0.003 & 0.69 \\
		NotFound$^2$ & 0.345 & -     & 0.137 & -     & 0.290 & -     & 0.129 & -     & 0.127 & - \\
		\hline
		Overall$^3$ &       & 35.74 &       & 31.67 &       & 34.16 &       & 29.94 &       & 29.93 \\
		\%DLT$^4$ & 12.9  &       & 16.4  &       & 13.6  &       & 17.6  &       & 17.4  &  \\
		\hline
		\multicolumn{11}{l}{$^\dagger$Frequency of being identified as the MTD. $^\ddagger$Average number of patients treated. } \\
		\multicolumn{11}{l}{$^1$All doses are deemed overly toxic. $^2$MTD cannot be declared after reaching maximum sample size.} \\
		\multicolumn{11}{l}{$^3$Average overall trial sample size. $^4$Observed \%DLT across all dose levels. }		
	\end{tabular}%
	\label{tab:16_33_sshaped}%
\end{table}%

\begin{table}[htbp]
	\centering
	\caption{Results: TTI (0.20, 0.30), S-shaped Curve}
	\begin{tabular}{c|cc|cc|cc|cc|cc}
		& \multicolumn{2}{c}{Original BLRM} & \multicolumn{2}{c}{Design 1} & \multicolumn{2}{c}{Design 2} & \multicolumn{2}{c}{Design 3} & \multicolumn{2}{c}{Design 4} \\
		MTD   & Frequency$^\dagger$ & N$^\ddagger$     & Frequency & N     & Frequency & N     & Frequency & N     & Frequency & N \\
		\hline
		AllToxic$^1$ & 0.021 & -     & 0.021 & -     & 0.021 & -     & 0.021 & -     & 0.021 & - \\
		10    & 0.000 & 3.00  & 0.000 & 3.00  & 0.000 & 3.00  & 0.000 & 3.00  & 0.000 & 3.00 \\
		25    & 0.000 & 3.10  & 0.000 & 3.03  & 0.000 & 3.06  & 0.000 & 2.96  & 0.000 & 2.96 \\
		50    & 0.000 & 4.44  & 0.000 & 3.50  & 0.000 & 3.95  & 0.000 & 3.03  & 0.000 & 3.05 \\
		100   & 0.002 & 18.59 & 0.007 & 10.09 & 0.004 & 14.27 & 0.008 & 8.47  & 0.010 & 8.93 \\
		\textbf{200} & \textbf{0.178} & \textbf{11.89} & \textbf{0.437} & \textbf{15.69} & \textbf{0.328} & \textbf{14.97} & \textbf{0.533} & \textbf{14.91} & \textbf{0.504} & \textbf{15.46} \\
		400   & 0.002 & 2.00  & 0.006 & 5.81  & 0.007 & 2.54  & 0.011 & 7.37  & 0.008 & 6.36 \\
		800   & 0.000 & 0.162 & 0.000 & 0.318 & 0.000 & 0.237 & 0.000 & 0.879 & 0.000 & 0.798 \\
		NotFound$^2$ & 0.796 & -     & 0.529 & -     & 0.639 & -     & 0.427 & -     & 0.457 & - \\
		\hline
		Overall$^3$ &       & 43.22 &       & 41.44 &       & 42.06 &       & 40.61 &       & 40.54 \\
		\%DLT$^4$ & 12.1  &       & 18.9  &       & 14.2  &       & 21.5  &       & 20.4  &  \\
		\hline
		\multicolumn{11}{l}{$^\dagger$Frequency of being identified as the MTD. $^\ddagger$Average number of pa	tients treated. } \\
		\multicolumn{11}{l}{$^1$All doses are deemed overly toxic. $^2$MTD cannot be declared after reaching maximum sample size.} \\
		\multicolumn{11}{l}{$^3$Average overall trial sample size. $^4$Observed \%DLT across all dose levels. }
	\end{tabular}%
	\label{tab:10_20_sshaped}%
\end{table}%

With a narrower TTI (0.20, 0.30), the maximum sample size 45 may be insufficient to identify the MTD with such high precision. Therefore, the accuracy of original BLRM can be poor, and all new designs are outperforming by a wider margin, especially for designs 3 and 4 (Table \ref{tab:10_20_sshaped}). Original BLRM tends to assign many patients to doses that are underdosing, and by the time it escalates to the true MTD, the sample size are almost exhausted. This issue is mitigated with new overdose control designs, especially in S-shaped and flat dose-toxicity scenarios. This is supported by the fact that the proportions of ``NotFound'' category are greatly reduced with new designs. 

\subsection{Simulation with Random Dose-Toxicity Scenarios}
The three fixed dose-toxicity scenarios in previous section are selected based on the shape of the curve, instead of dose-specific DLT rates. Simulation results have demonstrated encouraging improvement over the original overdose control strategy in BLRM. In this part of simulation, we apply each design 1,000 times and each time the underlying dose-toxicity scenario is randomly generated. This ensures the improvement observed is not due to cherry-picking of dose-toxicity scenarios. 

There are two popular algorithms to easily generate random monotonic dose-toxicity scenarios in a non-parametric manner: the pseudo-uniform algorithm\cite{ClertantOQuigley_2017} and the algorithm proposed by Paoletti and colleagues\cite{PaolettiEtal_2004}. We term them Clertant class and Paoletti class, respectively. Both classes require specifying a TTL $\phi$ which is set to 0.25 in the simulation. In addition to $\phi$, Clertant class only need the number of provisional doses, while Paoletti class has five parameters to control the deviation from $\phi$ at MTD and the variability at non-MTD dose levels. Details and examples of the two classes are provided in Appendix B. It is worth noting that neither algorithm is utilizing the actual dose strength. Indeed, the DLT rates can be generated based on the number of provisional dose levels first, then assigned to any sequence of dose levels in ascending order. It is conceivable that dose-toxicity scenarios generated this way are unlikely to be captured by the BLRM. Therefore, this part of simulation may be considered a stress test for the robustness of BLRM in the presence of model misspecification. 

Since MTD is random in each dose-toxicity scenario, the operating characteristics at each dose level are no longer useful. We hence only present the proportion out of 1,000 scenarios in which MTD is correctly identified (Table \ref{tab:sim_random}). In general, the performance of all designs are compromised with random dose-toxicity scenarios for the aforementioned reasons. The new designs are still able to outperform original BLRM, especially with TTI (0.16, 0.33). 

\begin{table}[htbp]
	\centering
	\caption{Accuracy of MTD Identification with Random Dose-Toxicity Scenarios}
	\begin{tabular}{c|cc|cc}
		\multirow{4}[0]{*}{Design} & \multicolumn{2}{c}{Clertant Class} & \multicolumn{2}{c}{Paoletti Class} \\
		& \multicolumn{2}{c}{TTI} & \multicolumn{2}{c}{TTI} \\
		& (0.20, 0.30) & (0.16, 0.33) & (0.20, 0.30) & (0.16, 0.33) \\
		& \%CorrectMTD & \%CorrectMTD & \%CorrectMTD & \%CorrectMTD \\
		\hline
		Original BLRM & 0.312 & 0.447 & 0.181 & 0.433 \\
		Design 1 & 0.329 & 0.522 & 0.248 & 0.482 \\
		Design 2 & 0.320 & 0.474 & 0.212 & 0.443 \\
		Design 3 & 0.348 & 0.535 & 0.259 & 0.517 \\
		Design 4 & 0.331 & 0.535 & 0.253 & 0.520 \\
	\end{tabular}%
	\label{tab:sim_random}%
\end{table}%

\subsection{Hypothetical Scenario Revisited}
We revisit the hypothetical data scenario in Table \ref{tab:data_scenario} to demonstrate how the new overdose control designs are implemented to avoid the issue presented. Suppose that the feasibility bounds $\alpha_f = 0.25$ in designs 1 and 3, and $g(r_i) = r_i$ in designs 2 and 4. Recall that the TTI is (0.16, 0.33) in the hypothetical data scenario and the current dose is at 100mg. BLRM with original overdose control at 0.25 is unable to make a recommendation to escalate to the next dose 200mg, as its associated overdose probability is 0.29. 

In design 1, the left-hand side (LHS) of \eqref{eqn:design1} can be calculated as $0.194 = 0.25\cdot 0.777$, whereas the right-hand side (RHS) is $0.028 = 0.75\cdot 0.037$. Therefore, the add-on rule is met and an escalation decision is rendered, as the LHS is larger than the RHS in \eqref{eqn:design1}. Original overdose control criterion is overridden in this case. In design 3, UPMs of underdosing and overdosing at current dose, $\mathcal{U}_{i}(Under)$ and $\mathcal{U}_{i}(Over)$, are respectively $4.856 = 0.777/0.16$ and $0.055 = 0.037/0.67$. Therefore, the add-on is also met with $\alpha_f = 0.25$. 

In design 2, the relative dose strength $r_i = 2$. The add-on rule is met because the LHS (0.777) is larger than the RHS (0.580 = $2\cdot 0.290$) in \eqref{eqn:design2}. Similarly, the add-on rule in UPM-based design 4 is also met as the LHS of the inequality in design 4 is 4.856, which is higher than the RHS $0.866 = 2\cdot 0.290/0.67$. 

\section{Concluding Remarks}
An accurately identified MTD serves as the cornerstone of successful subsequent phases in oncology drug development. BLRM is a useful model-based dose-finding approach because of its ability to utilize cumulative data and handle complicated dose-finding settings. It has been reported that BLRM with original overdose control strategy tends to be overly conservative. In this research, we compared the overdose control mechanisms in selected dose-finding designs and identified the reason that may have caused the conservativeness of BLRM. We proposed four alternatives via respective add-on rules in which the probability of underdosing is also accounted for. The scope of this manuscript is to improve the operating characteristics over original BLRM and our simulation results have demonstrated exactly that. It is of natural interest to compare the new designs with other popular dose-finding designs. This requires a comprehensive simulation study and relevant work is ongoing. 

In each new design, the add-on rule introduces an additional tuning parameter. The tuning parameters are intuitive and can be easily modified to achieve a desired balance between performance and conservativeness. For designs 1 and 3, the tuning parameter $\alpha_f$ has the same interpretation with the feasibility bound used in original EWOC method \cite{BabbEtal_1998CRMEWOC}, where a smaller $\alpha_f$ corresponds to a more conservative design. Performance may be improved if the data-dependent feasibility bound framework \cite{Wheeler_2017TDEWOC} can be extended here. For designs 2 and 4, the function form of $g(r_i)$ regulates how likely the add-on rule is triggered with the same relative dose strength. In our simulation, we consistently set $g(r_i) = r_i$. Alternatively, a power function $g(r_i) = r_i^c$ may be used where $c > 1$ results in a more conservative design, or vice versa. The feasibility bound $\alpha_f$ may also be introduced into designs 2 and 4. 

Conceptually, design 4 has the most appeal among new designs for two reasons. First, its add-on rule accounts for the relative dose strength $r_i$. Second, the add-on rule is UPM-based which factors in the length of the toxicity interval. Compared to its probability-base counterpart design 2 with the same $g(r_i)$, design 4 always has higher proportion of correct MTD identified and treats more patients at MTD. However, we notice that design 3 and 4 have almost identical operating characteristics, and they tend to assign more patients to dose levels that are higher than MTD. This observation seems to coincide with those made elsewhere \cite{YanEtal_2017Keyboard,ZhouEtal_2018DFCompare}. One plausible cause is the Occam's razor which states that more parsimonious models are favored in Bayesian inference. In designs 3 and 4, this means the underdosing interval, which is narrower than overdosing interval, will have a larger UPM. This could make the add-on rule get triggered more frequently than it should. In order to blunt the Occam's razor, mTPI-2 \cite{GuoEtal_2017mTPI2} and keyboard \cite{YanEtal_2017Keyboard} designs divided the underdosing/overdosing intervals into smaller intervals that have the same length with the target toxicity interval. Adopting a similar approach may improve the performance of designs 3 and 4 as well. 

Although the proposed new designs have shown better performance in the scenarios we tested, it is desirable to develop systematic approaches to trial design with certain optimality. The decision analytic framework fits this purpose. Using a decision analysis method, one can specify a utility index representing multiple objectives such as safety and efficacy \cite{HouedeEtal_2010}, hence. However, a remaining challenge is to find the dose escalation rule to maximize even a simple reward, e.g., the total number of subjects treated with the MTD. Although often there is no simple solution to this problem, one may find approximate solutions using, e.g., a hybrid design \cite{BartroffEtal_2010} or a multi-armed bandit approach, which has been used for dose-escalation trials only very recently \cite{AzizEtal_2021}. In general, we consider (approximate) optimal dose escalation designs as a promising approach for early drug development, although much research work is needed to develop the method and examine their performance in practical scenarios.

\bibliographystyle{wileyNJD-AMA}
\bibliography{references}

\newpage
\appendix
\section{Complete Fixed Scenario Simulation Results}
Tables A1 to A6 contain the complete simulation results in Section 3.1. 

%(0.16, 0.33), steep
\begin{table}[htbp]
	\centering
	\caption{Results: TTI (0.16, 0.33), Steep Curve}
	\begin{tabular}{c|cc|cc|cc|cc|cc}
		& \multicolumn{2}{c}{Original BLRM} & \multicolumn{2}{c}{Design 1} & \multicolumn{2}{c}{Design 2} & \multicolumn{2}{c}{Design 3} & \multicolumn{2}{c}{Design 4} \\
		MTD   & Frequency & N     & Frequency & N     & Frequency & N     & Frequency & N     & Frequency & N \\
		\hline
		AllToxic & 0.051 & -     & 0.051 & -     & 0.051 & -     & 0.051 & -     & 0.051 & - \\
		10    & 0.001 & 3.22  & 0.000 & 3.11  & 0.001 & 3.20  & 0.001 & 3.07  & 0.000 & 3.08 \\
		25    & 0.037 & 5.35  & 0.031 & 4.08  & 0.034 & 5.01  & 0.027 & 3.70  & 0.028 & 3.70 \\
		50    & 0.392 & 12.63 & 0.303 & 7.57  & 0.395 & 11.69 & 0.314 & 7.02  & 0.316 & 7.13 \\
		\textbf{100} & \textbf{0.318} & \textbf{6.54} & \textbf{0.470} & \textbf{8.80} & \textbf{0.354} & \textbf{6.54} & \textbf{0.462} & \textbf{7.91} & \textbf{0.456} & \textbf{8.07} \\
		200   & 0.012 & 0.76  & 0.036 & 2.52  & 0.019 & 1.09  & 0.049 & 3.55  & 0.049 & 3.38 \\
		400   & 0.000 & 0.07  & 0.001 & 0.19  & 0.000 & 0.10  & 0.000 & 0.56  & 0.000 & 0.53 \\
		800   & 0.000 & 0.00  & 0.000 & 0.00  & 0.000 & 0.00  & 0.000 & 0.05  & 0.000 & 0.04 \\
		NotFound & 0.188 & -     & 0.107 & -     & 0.145 & -     & 0.096 & -     & 0.100 & - \\
		\hline
		Overall &       & 28.60 &       & 26.30 &       & 27.65 &       & 25.85 &       & 25.93 \\
		\%DLT & 17.7 &       & 20.7 &       & 18.2 &       & 22.5 &       & 22.2 & \\
	\end{tabular}%
\end{table}%

%(0.16, 0.33), S-shaped
\begin{table}[htbp]
	\centering
	\caption{Results: TTI (0.16, 0.33), S-shaped Curve}
	\begin{tabular}{c|cc|cc|cc|cc|cc}
	& \multicolumn{2}{c}{Original BLRM} & \multicolumn{2}{c}{Design 1} & \multicolumn{2}{c}{Design 2} & \multicolumn{2}{c}{Design 3} & \multicolumn{2}{c}{Design 4} \\
	MTD   & Frequency & N     & Frequency & N     & Frequency & N     & Frequency & N     & Frequency & N \\
	\hline
	AllToxic & 0.017 & -     & 0.017 & -     & 0.017 & -     & 0.017 & -     & 0.017 & - \\
	10    & 0.000 & 3.01  & 0.000 & 3.01  & 0.000 & 3.01  & 0.000 & 3.00  & 0.000 & 3.00 \\
	25    & 0.000 & 3.07  & 0.000 & 3.03  & 0.000 & 3.07  & 0.000 & 2.97  & 0.000 & 2.97 \\
	50    & 0.000 & 3.74  & 0.000 & 3.11  & 0.000 & 3.70  & 0.001 & 3.02  & 0.001 & 3.03 \\
	100   & 0.047 & 14.30 & 0.051 & 7.91  & 0.045 & 12.08 & 0.060 & 6.62  & 0.057 & 6.70 \\
	\textbf{200} & \textbf{0.561} & \textbf{9.03} & \textbf{0.749} & \textbf{10.58} & \textbf{0.621} & \textbf{9.56} & \textbf{0.737} & \textbf{9.63} & \textbf{0.742} & \textbf{9.67} \\
	400   & 0.028 & 2.36  & 0.044 & 3.73  & 0.026 & 2.53  & 0.052 & 3.99  & 0.053 & 3.88 \\
	800   & 0.002 & 0.22  & 0.002 & 0.30  & 0.001 & 0.21  & 0.004 & 0.71  & 0.003 & 0.69 \\
	NotFound & 0.345 & -     & 0.137 & -     & 0.290 & -     & 0.129 & -     & 0.127 & - \\
	\hline
	Overall &       & 35.74 &       & 31.67 &       & 34.16 &       & 29.94 &       & 29.93 \\
	\%DLT & 12.9  &       & 16.4  &       & 13.6  &       & 17.6  &       & 17.4  &  \\
	\end{tabular}%
\end{table}%

%(0.16, 0.33), flat
\begin{table}[htbp]
	\centering
	\caption{Results: TTI (0.16, 0.33), Flat Curve}
	\begin{tabular}{c|cc|cc|cc|cc|cc}
	& \multicolumn{2}{c}{Original BLRM} & \multicolumn{2}{c}{Design 1} & \multicolumn{2}{c}{Design 2} & \multicolumn{2}{c}{Design 3} & \multicolumn{2}{c}{Design 4} \\
	MTD   & Frequency & N     & Frequency & N     & Frequency & N     & Frequency & N     & Frequency & N \\
	\hline
	AllToxic & 0.000 & -     & 0.000 & -     & 0.000 & -     & 0.000 & -     & 0.000 & - \\
	10    & 0.000 & 3.00  & 0.000 & 3.00  & 0.000 & 3.00  & 0.000 & 3.00  & 0.000 & 3.00 \\
	25    & 0.000 & 3.02  & 0.000 & 3.01  & 0.000 & 3.02  & 0.000 & 3.00  & 0.000 & 3.00 \\
	50    & 0.000 & 3.17  & 0.000 & 3.03  & 0.000 & 3.17  & 0.000 & 3.00  & 0.000 & 3.00 \\
	100   & 0.003 & 6.67  & 0.005 & 3.93  & 0.002 & 4.71  & 0.003 & 3.18  & 0.003 & 3.25 \\
	200   & 0.167 & 10.12 & 0.196 & 7.01  & 0.197 & 9.50  & 0.246 & 6.35  & 0.251 & 6.41 \\
	\textbf{400} & \textbf{0.524} & \textbf{7.76} & \textbf{0.685} & \textbf{8.87} & \textbf{0.595} & \textbf{8.51} & \textbf{0.624} & \textbf{7.93} & \textbf{0.615} & \textbf{7.92} \\
	800   & 0.039 & 2.48  & 0.042 & 3.24  & 0.030 & 2.44  & 0.040 & 3.77  & 0.043 & 3.73 \\
	NotFound & 0.267 & -     & 0.072 & -     & 0.176 & -     & 0.087 & -     & 0.088 & - \\
	\hline
	Overall &       & 36.20 &       & 32.09 &       & 34.34 &       & 30.23 &       & 30.31 \\
	\%DLT & 12.0  &       & 14.3  &       & 12.9  &       & 15.2  &       & 15.2  &  \\
	\end{tabular}%
\end{table}%

%(0.20, 0.30), steep
\begin{table}[htbp]
	\centering
	\caption{Results: TTI (0.20, 0.30), Steep Curve}
	\begin{tabular}{c|cc|cc|cc|cc|cc}
	& \multicolumn{2}{c}{Original BLRM} & \multicolumn{2}{c}{Design 1} & \multicolumn{2}{c}{Design 2} & \multicolumn{2}{c}{Design 3} & \multicolumn{2}{c}{Design 4} \\
	MTD   & Frequency & N     & Frequency & N     & Frequency & N     & Frequency & N     & Frequency & N \\
	\hline
	AllToxic & 0.066 & -     & 0.066 & -     & 0.066 & -     & 0.066 & -     & 0.066 & - \\
	10    & 0.000 & 3.20  & 0.000 & 3.07  & 0.000 & 3.17  & 0.000 & 3.05  & 0.000 & 3.06 \\
	25    & 0.004 & 7.73  & 0.001 & 4.89  & 0.003 & 6.09  & 0.002 & 3.92  & 0.002 & 4.08 \\
	50    & 0.157 & 21.14 & 0.150 & 12.73 & 0.186 & 18.59 & 0.162 & 10.70 & 0.172 & 11.22 \\
	\textbf{100} & \textbf{0.098} & \textbf{7.24} & \textbf{0.298} & \textbf{13.87} & \textbf{0.206} & \textbf{9.40} & \textbf{0.320} & \textbf{13.13} & \textbf{0.331} & \textbf{13.24} \\
	200   & 0.000 & 0.45  & 0.010 & 3.58  & 0.005 & 1.25  & 0.018 & 6.08  & 0.013 & 5.23 \\
	400   & 0.000 & 0.03  & 0.000 & 0.18  & 0.000 & 0.10  & 0.000 & 0.81  & 0.000 & 0.67 \\
	800   & 0.000 & 0     & 0.000 & 0.003 & 0.000 & 0.003 & 0.000 & 0.057 & 0.000 & 0.057 \\
	NotFound & 0.674 & -     & 0.475 & -     & 0.533 & -     & 0.432 & -     & 0.416 & - \\
	\hline
	Overall &       & 39.82 &       & 38.33 &       & 38.63 &       & 37.74 &       & 37.56 \\
	\%DLT & 16.8  &       & 22.1  &       & 18.7  &       & 24.8  &       & 24.1  &  \\
	\end{tabular}%
\end{table}%

%(0.20, 0.30), S-shaped
\begin{table}[htbp]
	\centering
	\caption{Results: TTI (0.20, 0.30), S-shaped Curve}
	\begin{tabular}{c|cc|cc|cc|cc|cc}
	& \multicolumn{2}{c}{Original BLRM} & \multicolumn{2}{c}{Design 1} & \multicolumn{2}{c}{Design 2} & \multicolumn{2}{c}{Design 3} & \multicolumn{2}{c}{Design 4} \\
	MTD   & Frequency & N     & Frequency & N     & Frequency & N     & Frequency & N     & Frequency & N \\
	\hline
	AllToxic & 0.021 & -     & 0.021 & -     & 0.021 & -     & 0.021 & -     & 0.021 & - \\
	10    & 0.000 & 3.00  & 0.000 & 3.00  & 0.000 & 3.00  & 0.000 & 3.00  & 0.000 & 3.00 \\
	25    & 0.000 & 3.10  & 0.000 & 3.03  & 0.000 & 3.06  & 0.000 & 2.96  & 0.000 & 2.96 \\
	50    & 0.000 & 4.44  & 0.000 & 3.50  & 0.000 & 3.95  & 0.000 & 3.03  & 0.000 & 3.05 \\
	100   & 0.002 & 18.59 & 0.007 & 10.09 & 0.004 & 14.27 & 0.008 & 8.47  & 0.010 & 8.93 \\
	\textbf{200} & \textbf{0.178} & \textbf{11.89} & \textbf{0.437} & \textbf{15.69} & \textbf{0.328} & \textbf{14.97} & \textbf{0.533} & \textbf{14.91} & \textbf{0.504} & \textbf{15.46} \\
	400   & 0.002 & 2.00  & 0.006 & 5.81  & 0.007 & 2.54  & 0.011 & 7.37  & 0.008 & 6.36 \\
	800   & 0.000 & 0.162 & 0.000 & 0.318 & 0.000 & 0.237 & 0.000 & 0.879 & 0.000 & 0.798 \\
	NotFound & 0.796 & -     & 0.529 & -     & 0.639 & -     & 0.427 & -     & 0.457 & - \\
	\hline
	Overall &       & 43.22 &       & 41.44 &       & 42.06 &       & 40.61 &       & 40.54 \\
	\%DLT & 12.1  &       & 18.9  &       & 14.2  &       & 21.5  &       & 20.4  &  \\
	\end{tabular}%
\end{table}%

%(0.20, 0.30), flat
\begin{table}[htbp]
	\centering
	\caption{Results: TTI (0.20, 0.30), Flat Curve}
	\begin{tabular}{c|cc|cc|cc|cc|cc}
	& \multicolumn{2}{c}{Original BLRM} & \multicolumn{2}{c}{Design 1} & \multicolumn{2}{c}{Design 2} & \multicolumn{2}{c}{Design 3} & \multicolumn{2}{c}{Design 4} \\
	MTD   & Frequency & N     & Frequency & N     & Frequency & N     & Frequency & N     & Frequency & N \\
	\hline
	AllToxic & 0.000 & -     & 0.000 & -     & 0.000 & -     & 0.000 & -     & 0.000 & - \\
	10    & 0.000 & 3.00  & 0.000 & 3.00  & 0.000 & 3.00  & 0.000 & 3.00  & 0.000 & 3.00 \\
	25    & 0.000 & 3.02  & 0.000 & 3.02  & 0.000 & 3.02  & 0.000 & 3.00  & 0.000 & 3.00 \\
	50    & 0.000 & 3.43  & 0.000 & 3.22  & 0.000 & 3.34  & 0.000 & 3.00  & 0.000 & 3.00 \\
	100   & 0.000 & 8.53  & 0.000 & 4.14  & 0.000 & 4.73  & 0.000 & 3.22  & 0.000 & 3.30 \\
	200   & 0.013 & 13.45 & 0.031 & 10.97 & 0.028 & 14.22 & 0.053 & 9.65  & 0.043 & 10.07 \\
	\textbf{400} & \textbf{0.164} & \textbf{10.35} & \textbf{0.364} & \textbf{13.86} & \textbf{0.280} & \textbf{12.25} & \textbf{0.422} & \textbf{13.30} & \textbf{0.433} & \textbf{13.62} \\
	800   & 0.001 & 2.373 & 0.001 & 4.635 & 0.001 & 2.7   & 0.004 & 6.381 & 0.004 & 5.661 \\
	NotFound & 0.821 & -     & 0.604 & -     & 0.691 & -     & 0.521 & -     & 0.520 & - \\
	\hline
	Overall &       & 44.19 &       & 42.84 &       & 43.25 &       & 41.54 &       & 41.65 \\
	\%DLT & 11.6  &       & 16.4  &       & 13.5  &       & 18.5  &       & 17.9  &  \\
	\end{tabular}%
\end{table}%

\section{Algorithms to Generate Random Dose-Toxicity Scenarios}
In this section, we present the details about two algorithms to easily generate random dose-toxicity curves for simulation purposes. 

\underline{Clertant Class}

This class is also known as the pseudo-uniform algorithm \cite{ClertantOQuigley_2017}. Only the number of dose levels $J$ and the target toxicity level $\phi$ is needed to implement this algorithm. Zhou and colleagues \cite{ZhouEtal_2018DFCompare_NoBLRM} provided a step-by-step instruction for this algorithm, which we paraphrase: 
\begin{enumerate}
	\item With equal probability, randomly pick a dose level out of $J$ as the MTD, denoted $j$.
	\item Randomly generate $M$ from $Beta(\max\{J-j, 0.5\}, 1)$ and let $B = \phi + (1-\phi)*M$.
	\item Repeatedly sample $J$ probabilities from $U(0, B)$ until these correspond to a scenario in which the $j$th dose is the MTD, i.e., closest to $\phi$. 
\end{enumerate} 

\underline{Paoletti Class}

The other algorithm is proposed by Paoletti and colleagues \cite{PaolettiEtal_2004}. This algorithm can be implemented in four steps \cite{LiuYuan_2015BOIN}: 
\begin{enumerate}
	\item With equal probability, randomly pick a dose level out of $J$ as the MTD, denoted $j$.
	\item Let $\Phi(\cdot)$ and $z(\cdot)$ be the CDF and inverse CDF of $N(0,1)$ respectively. Generate $p_j = \Phi(\epsilon_j)$, the DLT rate at MTD, where $\epsilon_j \sim N(z(\phi), \sigma_0^2)$.
	\item Generate $p_{j-1}$ and $p_{j+1}$ such that $p_j$ is closest to $\phi$. This guarantees dose $j$ is the MTD. 
	\begin{itemize}
		\item $p_{j-1} = \Phi[ z(p_j) - \{ z(p_j) - z(2\phi-p_j) \} \cdot I\{ z(p_j) > z(\phi) \} - \epsilon_{j-1}^2 ],$ where $\epsilon_{j-1} \sim N(\mu_1, \sigma_1^2)$
		\item $p_{j+1} = \Phi[ z(p_j) + \{ z(2\phi-p_j) - z(p_j) \} \cdot I\{ z(p_j) < z(\phi) \} + \epsilon_{j+1}^2 ],$ where $\epsilon_{j+1} \sim N(\mu_2, \sigma_2^2)$
	\end{itemize}
	\item Successively generate DLT rates for remaining dose levels. For $k \geq 2$,
	\begin{itemize}
		\item $p_{j-k} = \Phi\{ z(p_{j-k+1}) - \epsilon_{j-k}^2 \}$, where $\epsilon_{j-k} \sim N(\mu_1, \sigma_1^2)$
		\item $p_{j+k} = \Phi\{ z(p_{j+k-1}) - \epsilon_{j+k}^2 \}$, where $\epsilon_{j+k} \sim N(\mu_2, \sigma_2^2)$
	\end{itemize}
\end{enumerate} 

\begin{figure}
	\centering
	\includegraphics[width=\textwidth, height=\textheight, keepaspectratio]{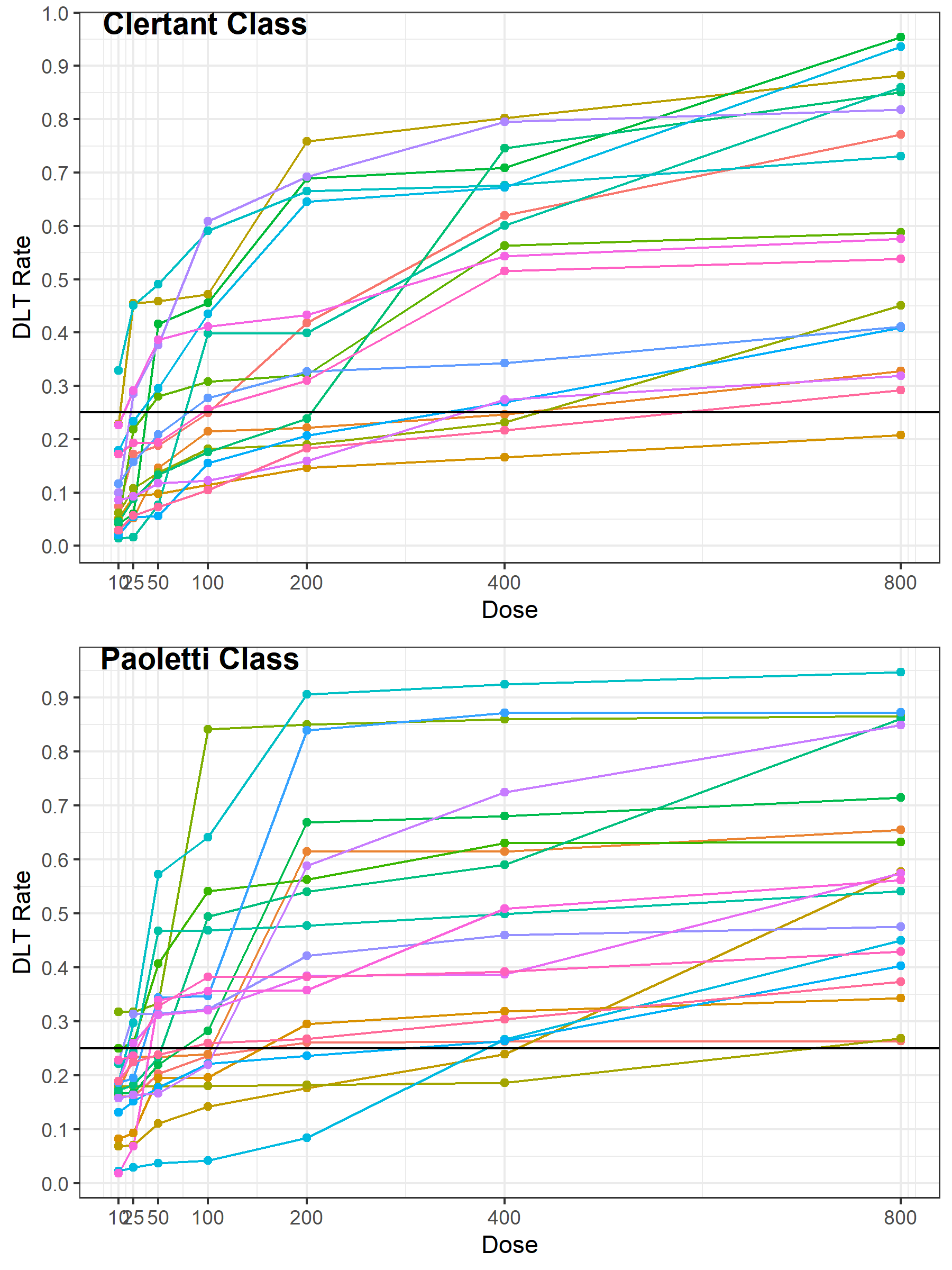}
	\caption{20 Random Scenarios By Each Class}
	\label{fig:random_scenarios}
\end{figure}

In addition to $\phi$, this algorithm has five tuning parameters $\{ \sigma_0, \mu_1, \sigma_1, \mu_2, \sigma_2 \}$. They are set to $\{ 0.1, 0.2, 0.3, 0.2, 0.4 \}$ in Section 3.2. Figure \ref{fig:random_scenarios} shows 20 randomly generated dose-toxicity scenarios by each class. The black horizontal line indicates TTL $\phi = 0.25$. 

\end{document}